# RECONFIGURABLE CONTROLLER DESIGN FOR ACTUATOR FAULTS IN A FOUR-TANK SYSTEM BENCHMARK

Pedram Hjiani[1] and Javad Poshtan[2*]

[1]Department of Electrical Engineering, Iran University of Science and Technology, Tehran, Iran

`p.hajiani@gmail.com`

[2]Department of Electrical Engineering, Iran University of Science and Technology, Tehran, Iran

`jposhtan@iust.ac.ir`

## ABSTRACT

*The purpose of this work is to design a state feedback controller using Parametric Eigenstructure Assignment (PAE) technique that has the capacity to be reconfigured in the case that partial actuator faults occur. The proposed controller is capable of compensating the gain losses in actuators and maintaining the control performance in faulty situations. Simulations show the performance enhancement in comparison to the non-reconfigurable controller through Integral Absolute Error (IAE) index for different fault scenarios.*

## 1. INTRODUCTION

Today's modern technological systems utilize sophisticated control systems to reach a high level of performance and security. In case of malfunctions in actuators, sensors, or other system components an ordinary feedback controller can make the system unstable, or its performance may degrade drastically [1]. Therefore, new control strategies have been developed that can automatically detect and compensate for system component faults, while maintaining the acceptable performance of the overall system. Such control systems are called Fault Tolerant Control Systems (FTCS) [1,2].

While a great deal of research work has done in the field of fault detection and diagnosis (FDD), much less effort has been devoted to fault tolerant control systems. Most of FDD techniques, however, are developed as a diagnosis or monitoring tool rather than an integral part of FTCS. As a result, some existing FDD methods may not satisfy the need of controller reconfiguration. On the other hand, most of the research on reconfigurable controls is carried out assuming the availability of a perfect FDD [1, 3].

---

[*]Corresponding Author
  

In this paper the focus is on Reconfigurable Controller (RC) that is a key part in fault tolerant control systems. The goal is to design a real time reconfigurable controller via eigenstructureassignment algorithm that compensates partial actuator failures in the system.The system under study is a MIMO nonlinear four-tank system benchmark. This system is a benchmark experimental facility developed for research purposes for process and aerospace industry [4-7].

Fault tolerant methods have been applied to multi-tank system benchmarks in a few recent research works. Some examples are mentioned in the following. In [8] fault tolerant has been implemented on a four-tank system using command governor (CG) controller. In [9] high order sliding mode observers have been used for a three-tank system. Authors of [10] have used predictive control and fuzzy logic to design a fault tolerant control for a three-tank benchmark. In [11] using feedback linearization, an approach has been proposed for fault tolerant control in a three-tank benchmark. Some other works in this respect are [12- 14].

The paper is organized as follows: In section 2 the model of the four tank system benchmark is described. The linearized model has been derived using perturbation theory. Section 3 is devoted to controller design method. Simulation results of implementation of the controller on the nonlinear model have been shown in section 4; comparing the performance of non reconfigurable controller with reconfigurable controller by some indices. And finally section 5 is conclusion and further works.

## 2. PROCESS DESCRIPTION

The process is called the Four-Tank system benchmark and consists of four interconnected water tanks and two pumps [15]. The system is shown in fig. 1. The inputs are the voltage to the two pumps that is at the standard range of 0-10 volts [16]. The outputs are the water levels in the lower tanks. The height of each tank is 20 cm. The four tank process can easily be built by using two double-tank processes, which are standard processes in many control laboratories[17,18].

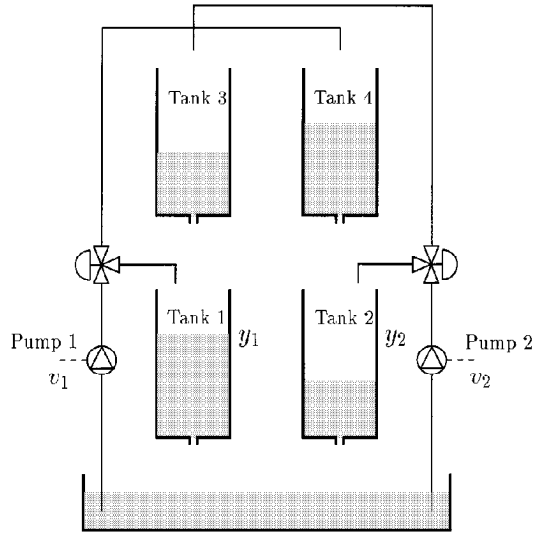

Figure 1. Schematic diagram of the four-tank process. The water levels in tanks 1 and 2 are controlled by two pumps.[15]

The target is to control the level in lower two tanks with two pumps. The process inputs are $v_1$ and $v_2$ (input voltage to the pumps) and the outputs are $y_1$ and $y_2$ (voltage from level measurement devices). Mass balance and Bernolli's law yields:[15]

$$\frac{dh_1}{dt} = -\frac{a_1}{A_1}\sqrt{2gh_1} + \frac{a_3}{A_1}\sqrt{2gh_3} + \frac{\gamma_1 k_1}{A_1}v_1$$

$$\frac{dh_2}{dt} = -\frac{a_2}{A_2}\sqrt{2gh_2} + \frac{a_4}{A_2}\sqrt{2gh_4} + \frac{\gamma_2 k_2}{A_2}v_2$$

$$\frac{dh_3}{dt} = -\frac{a_3}{A_3}\sqrt{2gh_3} + \frac{(1-\gamma_2)k_2}{A_3}v_2 \tag{1}$$

$$\frac{dh_4}{dt} = -\frac{a_4}{A_4}\sqrt{2gh_4} + \frac{(1-\gamma_1)k_1}{A_4}v_1$$

Where

$A_i$ cross-section of tank $i$;

$a_i$ cross-section of the outlet hole;

$h_i$ water level.

The voltage applied to pump $i$ is $v_i$ and the corresponding flow is $k_i v_i$. The parameters $\gamma_1, \gamma_2 \in (0,1)$ are determined from how the valves are set prior to an expriment. The flow to tank 1 is $\gamma_1 k_1 v_1$ and the flow to tank 4 is $(1-\gamma_1)k_1 v_1$ and similarly for tank 2 and tank 3. The acceleration of gravity is denoted g. The measured level signals are $k_c h_1$ and $k_c h_2$. The parameters values of the process are given in the following table:[15]

Table 1. process parameter values

| Parameter | Value |
|---|---|
| $A_1, A_3$ | $28\ cm^2$ |
| $A_2, A_4$ | $32\ cm^2$ |
| $a_1, a_3$ | $0.071\ cm^2$ |
| $a_2, a_4$ | $0.057\ cm^2$ |
| $k_c$ | $0.50\ V/cm$ |
| $g$ | $981\ cm/s^2$ |

Introduce the variables $x_i \triangleq h_i - h_i^0$ and $u_i \triangleq v_i - v_i^0$. The linearized state space equation is then given by

$$\frac{dx}{dt} = \begin{bmatrix} -\frac{1}{T_1} & 0 & \frac{A_3}{A_1 T_3} & 0 \\ 0 & -\frac{1}{T_2} & 0 & \frac{A_4}{A_2 T_4} \\ 0 & 0 & -\frac{1}{T_3} & 0 \\ 0 & 0 & 0 & -\frac{1}{T_4} \end{bmatrix} x + \begin{bmatrix} \frac{\gamma_1 k_1}{A_1} & 0 \\ 0 & \frac{\gamma_2 k_2}{A_2} \\ 0 & \frac{(1-\gamma_2) k_2}{A_3} \\ \frac{(1-\gamma_1) k_1}{A_4} & 0 \end{bmatrix} u \ , \ y = \begin{bmatrix} k_c & 0 & 0 & 0 \\ 0 & k_c & 0 & 0 \end{bmatrix} x \quad (2)$$

Where the time constants are:

$$T_i = \frac{A_i}{a_i} \sqrt{\frac{2 h_i^0}{g}} \quad i = 1,...,4 \qquad (3)$$

The operating point parameters are shown in table 2 [15].

Table 2. Operating point parameter values of the process

| Parameter | Value |
|---|---|
| $(h_1^0, h_2^0, h_3^0, h_4^0)$ | $(12.4, 12.7, 1.8, 1.4)[cm]$ |
| $(v_1^0, v_2^0)$ | $(3.00, 3.00)[V]$ |
| $(k_1, k_2)$ | $(3.33, 3.35)[cm^3/Vs]$ |
| $(\gamma_1, \gamma_2)$ | $(0.70, 0.60)$ |

Substituting operating point parameters in equations 2 and 3 the state space form is as follows:

$$\dot{x} = \begin{bmatrix} -0.0159 & 0 & 0.0419 & 0 \\ 0 & -0.0111 & 0 & 0.0333 \\ 0 & 0 & -0.0419 & 0 \\ 0 & 0 & 0 & -0.0333 \end{bmatrix} x + \begin{bmatrix} 0.0833 & 0 \\ 0 & 0.0628 \\ 0 & 0.0479 \\ 0.0312 & 0 \end{bmatrix} u \ , \ y = \begin{bmatrix} 0.5 & 0 & 0 & 0 \\ 0 & 0.5 & 0 & 0 \end{bmatrix} x \quad (4)$$

The states can be measured directly through tanks levels so there is no need for designing observer.

## 3. CONTROLLER DESIGN

The system is observable and controlabe. The states of the system are available through direct measurement of tanks levels. The process state and output equations are considered as the following:

$$\dot{x} = Ax + Bu$$
$$y = Cx \qquad (5)$$

International Journal of Instrumentation and Control Systems (IJICS) Vol.2, No.2, April 2012

Where $x_{(t)} \epsilon R^n$ is the state vector, $u_{(t)} \epsilon R^m$ is the input vector and $y_{(t)} \epsilon R^p$ is the output vector. A, B and C are system, input and output matrices respectively. By defining $e_{(t)} = r_{(t)} - y_{(t)}$ and augment the states e with system states x we can have the integral action in the controller for better tracking. The augmented system equations are as the following:[19]

$$\dot{\tilde{x}} = \tilde{A}\tilde{x} + \tilde{B}\tilde{u}$$
$$y = Cx$$
$$u = -\begin{bmatrix} k_1 & k_2 \end{bmatrix}\tilde{x}$$

where $\tilde{x} = \begin{bmatrix} x \\ e \end{bmatrix}$, $\tilde{A} = \begin{bmatrix} A & 0 \\ -C & 0 \end{bmatrix}$, $\tilde{B} = \begin{bmatrix} B \\ 0 \end{bmatrix}$ (6)

Using Ackerman's method we can calculate the gain of state feedback controller.[20] In this paper the poles are selected as $p = [-0.252 \ -0.184 \ -0.017 \ -0.057 \ -0.073]$. For the reconfigurable controller the gain matrix is calculated online based on the real time values of system and input matrices.

## 4. SIMULATION RESULTS

The system and controller is simulated in Matlab Simulink environment. The system is once controlled with a nonreconfigurable state feedback controller. At times 100s and 350s step changes in the reference signal have occurred. At time 200s actuators gains has fallen by 60 percent suddenly. With this fault in the system there is a step change of reference signal in time 350s. All this scenario has been repeated using the reconfigurable controller. It is assumed that a proper FDI module is present and detects the fault after a delay time of 1s. Fig. 2 shows the actuator gain during simulation time. Fig. 3 and 4 show the comparison of the system outputs under the two control strategies. These figures illustrate the better performance of the reconfigurable controller and it has compensated for the actuator faults occurred in the system.

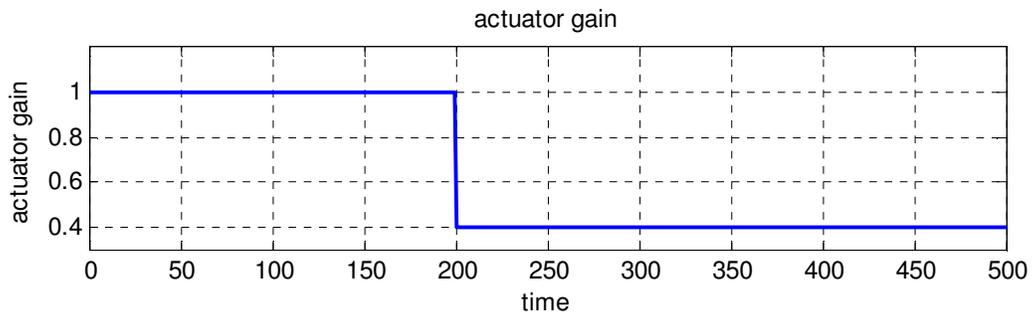

Figure 2. Actuators gains

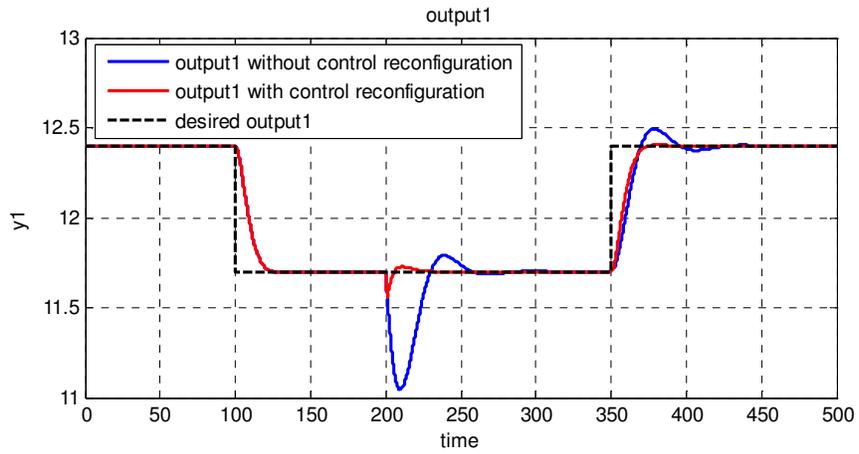

Figure 3. Output 1 of process with and without controller reconfiguration

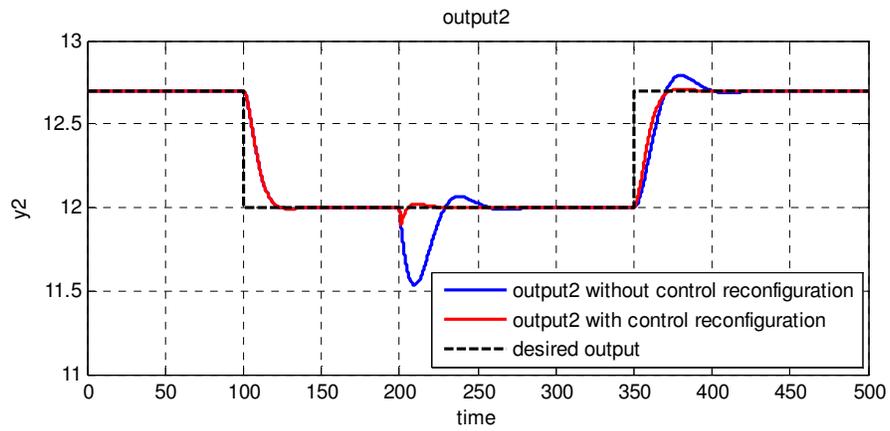

Figure 4. Output 2 of process with and without controller reconfiguration

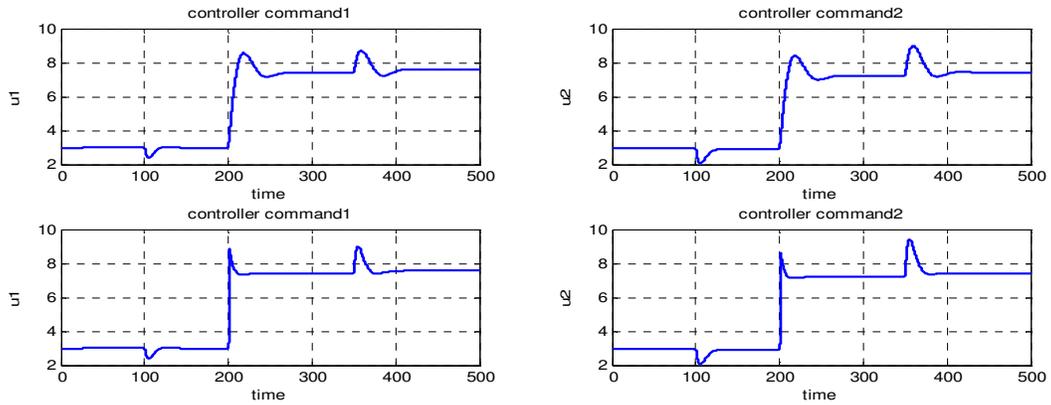

Figure 5. Controller command signals. Top: Without controller reconfiguration Bottom: With controller reconfiguration

In table 3 the performance of the two controllers are compared using the Integral Absolute Error (IAE) index. The IAE index is calculated while the actuator gain has fallen, once for the step change and the other time for the transient response caused by sudden change of the actuators gain.

Table 3. Comparison of controllers quality by IAE factor

| IAE factor | Without controller reconfiguration | | With controller reconfiguration | |
| --- | --- | --- | --- | --- |
| | Output1 | Output2 | Output1 | Output2 |
| Transient response of actuator gain fall | 13.07 | 9.231 | 0.702 | 0.501 |
| Step change in reference signal | 9.348 | 9.405 | 6.189 | 6.234 |

## 2.9. Acknowledgements

An (unnumbered) acknowledgements section may be inserted if required.

## 2.10. References

References should be cited in the main text, in passing [1] or explicitly as in [2]. The full references should be given as below (essentially IEEE format), in the order in which they are cited, in 10 pt. Times New Roman, with a 6pt spacing between each.

## 3. CONCLUSIONS

A tracking reconfigurable state feedback controller using Eigenstructure assignment method has been designed for a Four-Tank system benchmark and it is simulated with nonlinear model using Matlab software. The results show good performance of the controller and its ability to compensate for faults in actuators; that is a decrease in actuator gains by 60 percent. The performance is compared with the case that the controller state feedback gain is constant using IAE index. It has been shown that the reconfigurable controller has a significant merit on the non-reconfigurable controller.

This controller is proved to have capability of compensating actuator faults in real time. We are going to integrate this control with fault detection and isolation methods to achieve a fault tolerant control that automatically detects faults in actuators and compensate for them. In future works this method is going to be tested on a real laboratory test bed.